\documentclass[12pt]{article}
\usepackage[margin=1in]{geometry}
\usepackage{amssymb,amsmath}
\usepackage{graphicx}
\usepackage{subcaption}
\usepackage{enumitem}
\usepackage{xcolor}
\usepackage{setspace}
\usepackage[round]{natbib}
\usepackage{algorithm, algorithmic}

\newcommand{\x}{\mathbf{X}}
\newcommand{\y}{\mathbf{Y}}
\newcommand{\z}{\mathbf{Z}}
\newcommand{\w}{\mathbf{W}}

\newcommand{\bt}{\boldsymbol{\beta}}
\newcommand{\et}{\boldsymbol{\eta}}
\newcommand{\tht}{\boldsymbol{\theta}}
\newcommand{\dt}{\boldsymbol{\Delta}}

\begin{document}

\title{Log-Gaussian Cox Process Modeling of Large Spatial Lightning Data using Spectral and Laplace Approximations}
\author{Megan L. Gelsinger; Maryclare Griffin; David S. Matteson; Joseph Guinness}
\date{\today}

\maketitle

\begin{abstract}
    \noindent Lightning is a destructive and highly visible product of
    severe storms, yet there is still much to be learned about the
    conditions under which lightning is most likely to occur.  The
    GOES-16 and GOES-17 satellites, launched in 2016 and 2018 by NOAA
    and NASA, collect a wealth of data regarding individual lightning
    strike occurrence and potentially related atmospheric
    variables. The acute nature and inherent spatial correlation in
    lightning data renders standard regression analyses
    inappropriate. Further, computational considerations are
    foregrounded by the desire to analyze the immense and rapidly
    increasing volume of lightning data. We present a new
    computationally feasible method that combines spectral and Laplace
    approximations in an EM algorithm, denoted SLEM, to fit the widely
    popular log-Gaussian Cox process model to large spatial point
    pattern datasets.  In simulations, we find SLEM is competitive
    with contemporary techniques in terms of speed and accuracy. When
    applied to two lightning datasets, SLEM provides better
    out-of-sample prediction scores and quicker runtimes, suggesting
    its particular usefulness for analyzing lightning data, which tend
    to have sparse signals.
\end{abstract}


\section{Introduction}\label{intro}

Lightning has great destructive capabilities, and there is growing concern surrounding the relationship between climate change and lightning activity \citep{clark2017parameterization,finney2018projected}.  In 2016, lightning was added to the Global Climate Observing System’s (GCOS) list of Essential Climate Variables, indicators of particular focus for scientists looking to understand and mitigate climate impacts~\citep{GCOS2016}.  In October 2017, scientists associated with GCOS and several other meteorological organizations assembled a task force to spearhead a new wave of lightning research \citep{aich2018lightning}.  
Satellite data was identified as a crucial source of information for future lightning study \citep{aich2018lightning}. 
Modern satellite technology is capable of monitoring lightning activity over large, e.g. 1000$\times$1000 km, spatial grids.  This spatial scale allows researchers to conduct novel studies of  macro-level lightning dynamics, but poses a challenge to computational feasibility.  Another difficulty lies in the sparsity of lightning count data.  At most locations at any given time, there is no lightning, which means that the vast majority of recorded counts are zero. These problems motivate our study of computationally feasible statistical methods for satellite lightning data.

\begin{figure}
\begin{center}
\includegraphics[width = \textwidth]{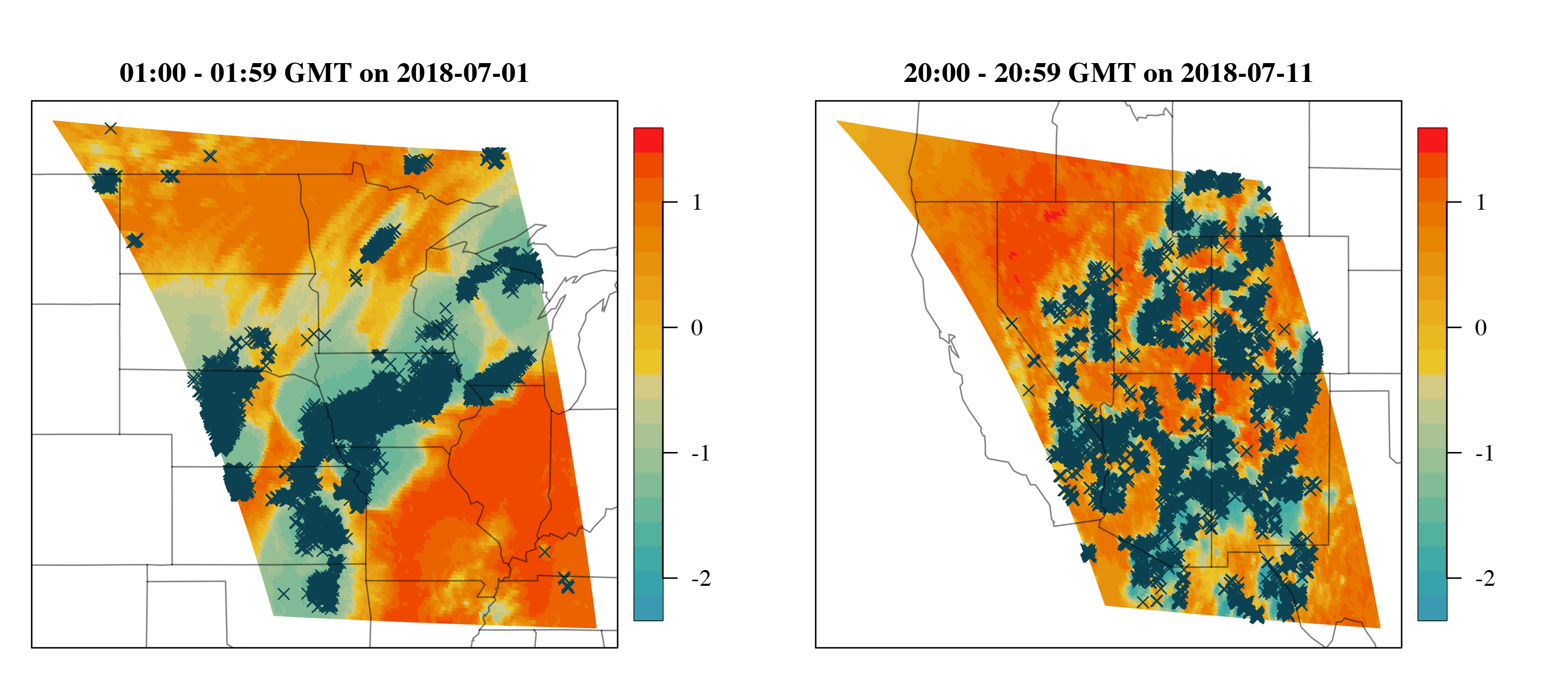}
\end{center}
\caption{Lightning strike locations $\left(\times\right)$  overlaid on proxy data for cloud-top height.}
\label{fig:real1_wABI}
\end{figure}

We concentrate on data collected by instruments on the GOES-16 satellite, launched in 2016 by the National Oceanic and Atmospheric Administration (NOAA) and the National Aeronautics and Space Administration (NASA).  The first instrument of interest to our study is the Advanced Baseline Imager (ABI), which records images in sixteen different spectral bands, corresponding to environmental factors such as water-based cloud coverage and dust, haze, and smoke presence.  We focus on the ABI's mesoscale mode of operation, which collects information over approximately a 1000 x 1000 km field-of-view, often targeted at areas of intense storm activity in North America.  This field-of-view can change hourly as the pattern of storms changes. Data is recorded on a minute-by-minute basis, at either 2 km, 1 km, or 0.5 km resolution, depending on the spectral band \citep{noaa2018goes}. The second instrument is the Geostationary Lightning Mapper (GLM), which continuously measures all types of lightning activity at an 8 km resolution over the Americas and adjacent oceanic regions.  Flashes are detected by their radiance signature--optical pulses which exceed the background instrument threshold \citep{goodman2013goes}.  For ease of comparison, we analyze both the ABI and GLM data at an 8 km spatial resolution.  
Figure \ref{fig:real1_wABI} shows environmental proxies derived from ABI data overlaid with GLM strike data. Construction of these environmental proxies is later described in detail in Section \ref{cov_descr}.

Log-Gaussian Cox process (LGCP) models are commonly used to model spatial point pattern data like the lightning strikes in Figure~\ref{fig:real1_wABI} \citep{moller1998log}.  To define the LGCP model, consider a point pattern whose locations $\mathbf{W} = \{W_1,...,W_{\ell}\}$ fall within the domain $S\subset\mathbb{R}^2$. As is common when analyzing point patterns, we use a discretization approximation and partition the domain $S$ into a regular $n_1\times n_2$ grid with $n$ pixels $\{B_1,\hdots,B_n\}$ and pixel centroids $\{s_1,\hdots,s_n\}$.  We transform the observed locations into counts per pixel via $Y_i = \sum_{j = 1}^{\ell}\mathbf{1}_{W_j\in B_i}$, for $i=1,\hdots,n$.  We also consider $\x(s_i)$, a $1\times (p + 1)$ row vector of an intercept and covariates considered constant within the $i$th pixel. An LGCP model for $\mathbf{W}$ implies the following model for $Y_1,\ldots,Y_n$:
\begin{align}
Y_i|\lambda &\stackrel{indep}{\sim} \textup{Poisson}\left(\int_{B_i}\lambda(s) \, ds\right) = \textup{Poisson}\!\left\{\Delta_i\lambda(s_i)\right\}\label{eq:poiss_part}\\
\lambda(s_i) &= \exp\{\x(s_i)\bt + Z(s_i)\}\label{eq:int_part}\\
Z(s)&~\sim GP\{0, K(\boldsymbol \eta)\label{eq:gp_part}\}
\end{align}
where $\lambda$ is an intensity function that is constant within each pixel, $\Delta_i$ is the area of pixel $B_i$, $\bt$ is a $(p + 1)\times 1$ vector of coefficients, and $K(\boldsymbol \eta)$ is a covariance function parameterized by $\et$.   Letting $\z = \left(Z(s_1),\dots, Z(s_n)\right)$ denote points from the Gaussian field, this is equivalent to assuming that $\z \sim N\left(\boldsymbol 0, \Sigma_{\et}\right)$. Even with the discretezation approximation, evaluating the likelihood remains challenging due to integration over the random effects $\z$:
\begin{align}
L(\tht; \y) = p(\y|\tht) = \int_{\mathbb{R}^n} p(\y,\z|\tht) \, d\z = \int_{\mathbb{R}^n} p(\y|\z,\tht)p(\z|\tht) \, d\z, \label{eq:like}
\end{align}
where $\y = (Y_1,\hdots,Y_n)$ and $\tht = (\bt, \et)$.

Markov Chain Monte Carlo (MCMC) methods are popular for fitting Bayesian LGCP models, providing exact inference given infinitely many samples from the posterior.  \cite{brix2001spatiotemporal} and \cite{diggle2005point} implement the Metropolis-adjusted Langevin kernel discussed in \cite{moller1998log} in an MCMC routine for LGCPs, making use of circulant embedding \citep{wood1994simulation}, which leverages fast Fourier transforms to speed up matrix computations.
This method is implemented in the R package \texttt{lgcp}~\citep{taylor2013lgcp, taylor2015bayesian}.  While the ``exactness" of this method is appealing, it is also known to have slow runtime, can mix poorly, and requires specification of user-defined tuning parameters \citep{taylor2014inla, shirota2016inference}. To address these issues, \cite{guan2018computationally}, introduce an approximate method which projects the random effects onto a lower-dimensional subspace. This reduces the dimension of the random effects and alleviates spatial confounding. Likewise, sampling the random effects involves manipulation of a lower dimensional matrix with better mixing properties.

Maximum likelihood schemes are also popular, but approximations are used due to the intractability of evaluating the likelihood. \cite{guan2020fast} use an Expectation-Maximization (EM) algorithm in which the E-step is approximated via sampling or Laplace approximation.  \cite{park2020reduced} use a Monte Carlo likelihood approximation instead, introducing a method for finding a good importance function iteratively. Both \cite{guan2020fast} and \cite{park2020reduced} also use similar projection-based approximations to \cite{guan2018computationally} to reduce computational burden and address spatial confounding. 

An especially well-known  approximation method is the integrated nested Laplace approximation (INLA).  As the name suggests, the key feature of INLA is its nested approximation of the marginal posterior distribution of the model's hyperparameters, such as $\et$, via the Laplace approximation \citep{rue2009approximate, illian2012toolbox}. INLA assumes that the Gaussian process driving the spatial point process is a Gaussian Markov random field, and thus has sparse precision matrices \citep{lindgren2011explicit},  facilitating faster matrix operations.  

In a related alternative, \cite{zilber2021vecchia} combine the Laplace approximation with a computationally efficient Vecchia approximation to the latent Gaussian process, implemented in the R package GPvecchia \citep{GPvecchia}. \cite{guan2020fast} also propose a variant to their method which leverages the Laplace approximation in the E-step, instead of using Monte Carlo averages.  
However, scalability to datasets measured on large spatial grids, on the order of tens of thousands as opposed to hundreds of locations, still remains in question even with the general computational time advantages of these methods compared to MCMC based approches \citep{taylor2014inla, guan2020fast}.

Despite recent advances, computational considerations remain critical due to the ever-increasing sizes of modern datasets. 
In this work, (i) we introduce an EM algorithm which leverages both the Laplace approximation and fast and scalable FFT algorithms to facilitate matrix computations. 
While spectral methods are powerful, they do not solve all of the computational challenges within the EM algorithm. (ii) To address these remaining challenges, we also use the Hutchinson trace approximation  \citep{hutchinson1989stochastic}.  (iii) Additionally, we craft a local covariance matrix approximation that can be combined with the Laplace approximation to approximate  the conditional mean of the Gaussian field given the data, after $\et$ and $\bt$ have been estimated. 
Combined, these techniques form the proposed Spectral-Laplace-Expectation-Maximization (SLEM) method for efficient estimation of LGCP models from large spatial point pattern data. In simulations, SLEM yields sizeable computatational advantages, with faster runtimes than the Vecchia-Laplace method. 
These runtime gains are accompanied by competitive estimation of $\boldsymbol \beta$ and come at the cost of less accurate estimation of the latent field. On the lightning datasets we consider, which are sparser than the simulation data, SLEM is both faster and more accurate on an out-of-sample log score prediction metric.

\subsection{Expository Analysis of Lightning Data}\label{prelim_light}
To illustrate the difficulty of estimating LGCPs on GOES lightning data, which have both large spatial scale (125$\times$125 pixel grid) and sparse signals (few pixels with non-zero strikes), we present results from two contemporary techniques, the Vecchia-Laplace (VL) algorithm, as implemented in the R package \texttt{GPvecchia} \citep{GPvecchia}, and INLA, as implemented in the \texttt{INLA} and \texttt{inlabru} R packages \citep{martins2013bayesian, bachl2019inlabru}.

\begin{figure}
\begin{center}
\includegraphics[width = \textwidth]{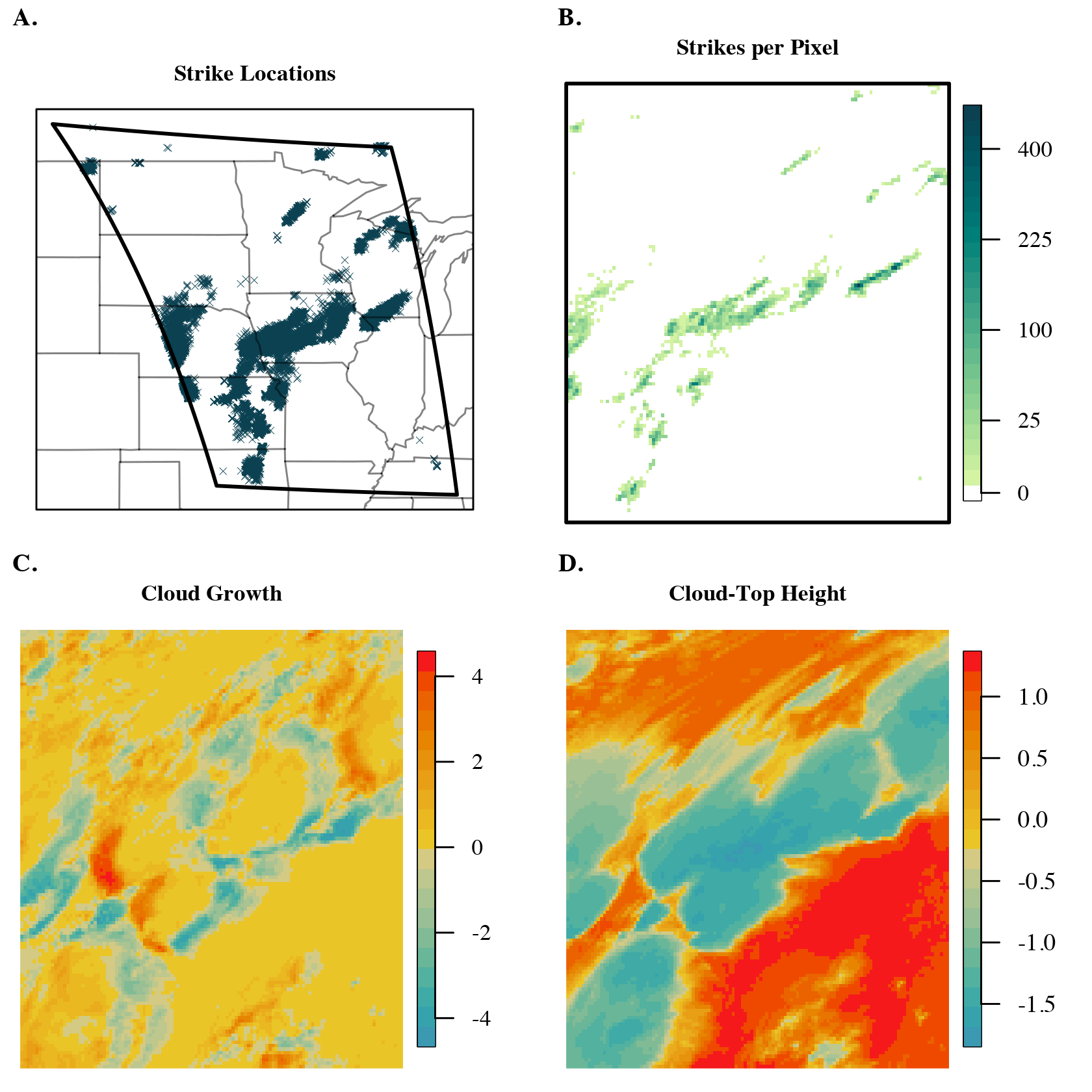}
\end{center}
\caption{\textbf{A.} Individual lightning strikes recorded from 01:00 - 01:59 GMT on 2018-07-01 in designated area of North America.  \textbf{B.} Lightning strikes from A. converted to counts per pixel on 125$\times$125 grid, colors on square-root scale. \textbf{C , D.}  Cloud growth and cloud-top height proxy data, respectively, averaged over same time frame as A and grid as B. Covariates are centered and scaled.}
\label{fig:real1}
\end{figure}

We apply both methods to GLM lightning data collected in a designated area of North America between 01:00 - 01:59 GMT on 2018-07-01. We convert the strikes to counts per pixel on an evenly spaced 125$\times$125 grid ($n = 15,\!625$) in order to model the lightning and covariate data on the same spatial scale. We include an intercept and several covariates in the model. The covariates include proxies for cloud growth and cloud-top height, which are later described in Section \ref{cov_descr}.  We also use elevation as a covariate in our model.  These covariates are currently believed to be associated with lightning occurrence, making them natural predictors to include in the model \citep{henderson2021evaluating, lee2021simplified, kilinc2007spatial, kotroni2008lightning}. All covariates are centered and scaled before including them in the model.  Their inclusion means that we interpret the Gaussian field as the effect of environmental factors on lightning intensity after controlling for cloud growth, cloud-top height, and elevation information. Figure \ref{fig:real1} provides visuals of the lightning strikes and several covariates.

\begin{figure}
\begin{center}
\includegraphics[width = \textwidth]{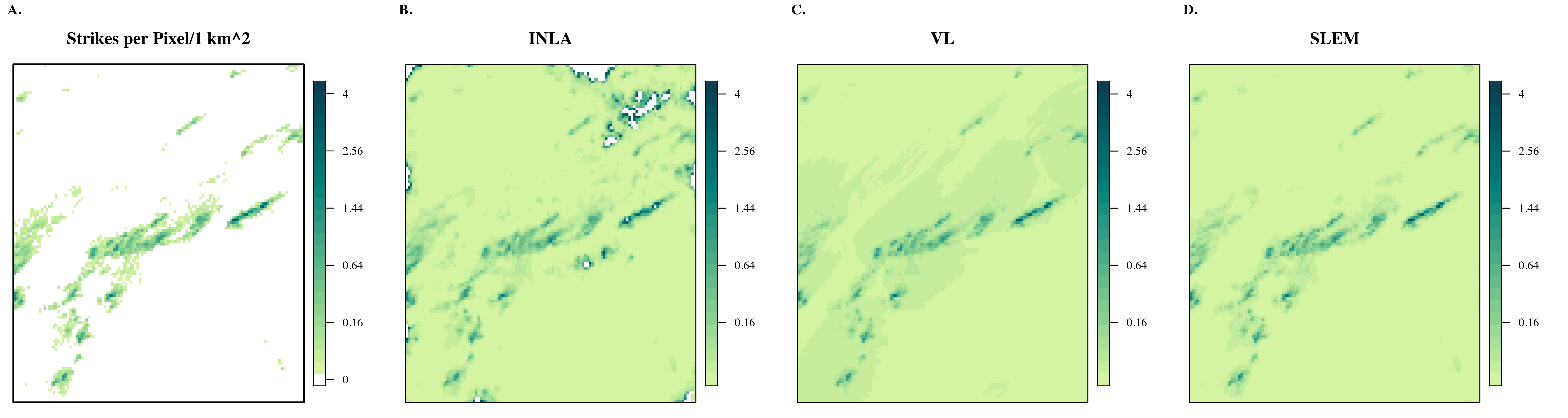}
\end{center}
\caption{Counts and intensity per 1 km$^2$/hour estimates returned by INLA (153 min. CPU time), VL (1865 min.), and SLEM (207 min.) for lightning dataset 1 shown in Figure~\ref{fig:real1}.  We employ a square-root transformation for better image definition.  White spaces in B and C reflect estimates exceeding indicated range.}
\label{fig:real1_res_comp}
\end{figure}

Figure \ref{fig:real1_res_comp} shows the estimated intensity per 1 km$^2$/hour returned by INLA, VL, and SLEM.  INLA appears to estimate a finer scale of lightning activity than is actually present in the observed counts.  The spurious locations of activity are especially troubling as they seem associated with relatively large intensity values, as evidenced by the white spaces in Figure \ref{fig:real1_res_comp} B which indicate values exceeding the plotting range.  Not only does INLA estimate activity where it is not present, it estimates a large amount of activity there.  The VL algorithm returns possibly overly smooth estimates of the intensity, washing out isolated areas of activity and grouping more closely occurring ones together. The proposed SLEM approach provides a middle-ground between INLA and VL.  SLEM's estimated intensity captures fine scale lightning patterns like INLA, but without introducing the same spurious activity.  SLEM also provides recognition of larger areas of activity like VL, but does so with less smoothing.  

In the sections that follow, we define the SLEM method and describe each of its components in detail.  We then perform in-depth studies of SLEM and VL applied to simulated and lightning datasets, focusing on VL as the most competitive method based on preliminary results.

\section{Methodology}\label{method}

\subsection{EM applied to LGCP}\label{slem}

Throughout our analysis, and to improve convergence of our algorithm below, we perform a change of variables
\begin{align}
W(s) = \x(s)\bt + Z(s),
\end{align}
which results in the likelihood function
\[L(\tht; \y) = p(\y|\tht) = \int p(\y,\w|\tht) \, d\w = \int p(\y|\w,\tht)p(\w|\tht) \, d\w.\]
We obtain a value $\tht$ that approximately maximizes the likelihood function iteratively using an approximate EM algorithm. Let $\tht^{(t)}$ refer to the value of the parameters at iteration $t$. At iteration $t+1$, the EM algorithm finds a new value $\tht^{(t+1)}$ by increasing the objective function, 
\begin{align}
Q(\tht|\tht^{(t)})
 &= E_{\w|\y,\tht^{(t)}}\left[  \,\log\{p(\y|\w,\tht)\}  \,\right] + E_{\w|\y,\tht^{(t)}}\left[  \,\log\{p(\w|\tht)\}  \,\right]\nonumber\\
&= E_{\w|\y,\tht^{(t)}}\left[  \,\sum_{j=1}^nY_j[\log(\Delta_j) + W_j] - \Delta_j\exp(W_j)-\log(Y_j!)  \,\right]\nonumber\\
&+ E_{\w|\y,\tht^{(t)}}\left[ \,-\frac{1}{2}\left[\log(|\Sigma_{\et}|)+(\w - X\bt)^{T}\Sigma^{-1}_{\et}(\w - X\bt)+n\log(2\pi) \,\right]\right]\nonumber .
\end{align}
In our alternative parameterization, $E_{\w|\y,\tht^{(t)}}(W_j)$ and $E_{\w|\y,\tht^{(t)}}(\exp(W_j))$ depend on $\tht^{(t)}$ but not $\tht$, so the objective function simplifies to
\begin{align}
Q(\tht|\tht^{(t)})
=& -\frac{1}{2}\left[\log(|\Sigma_{\et}|)+ 
(E_{\w|\y,\tht^{(t)}}[\,\w \,] - X\bt)^T\Sigma^{-1}_{\et}(E_{\w|\y,\tht^{(t)}}[\,\w \,] - X\bt)\right.\nonumber\\ 
&+\left.
\textup{tr}\left(\Sigma^{-1}_{\et}E_{\w|\y,\tht^{(t)}}[\, (\w - E_{\w|\y,\tht^{(t)}}[\,\w \,])(\w - E_{\w|\y,\tht^{(t)}}[\,\w \,])^T\,]\right)\right] + c,\nonumber
\end{align}
where $c$ contains terms that do not depend on $\tht$.
Evaluating the objective function is computationally challenging and requires a novel approach. The next several subsections detail how we perform computations and approximate this objective function in SLEM.

\subsection{Circulant Covariance Assumption}\label{model}

We assume that $K(\boldsymbol \eta)$ is the circulant version of the quasi-Mat\'ern covariance function with variance and range parameters $\sigma$ and $\alpha$, as presented in \cite{guinness2017Circulant}. 
Letting $\boldsymbol \omega$ refer to the Fourier frequencies associated with the spatial grid, the quasi-Mat\'ern covariance function is defined as 
\[
\mbox{Cov}(Z(\boldsymbol{s}),Z(\boldsymbol{s}+\boldsymbol{h})) = \frac{1}{n}\sum_{\boldsymbol{\omega}} \left(\sigma^2\left(1 + \alpha^2\sin^2\left(\frac{\omega_1}{2}\right) + \alpha^2\sin^2\left(\frac{\omega_2}{2}\right)\right)^{-2}\right) e^{i \boldsymbol{\omega} \cdot \boldsymbol{h}} d \boldsymbol{\omega}.\]
Given this spectral representation  $\et = (\sigma^2, \alpha)$ and $\Sigma_{\et}$ is block circulant.
The circulant assumption on $\Sigma_{\et}$ simplifies evaluation of $\log(|\Sigma_{\et}|)$, as the log determinant of a circulant matrix is equal to the sum of the log spectral density evaluated at the Fourier frequencies. The circulant assumption also allows for fast matrix-vector multiplications $\Sigma_{\et}^{-1} \boldsymbol{v}$, which are leveraged for evaluating other terms in the objective function.

\subsection{Laplace Approximation}\label{subsubsec:lapapp}

As described in Section \ref{intro}, it is common to approximate the distribution $p(\w|\y,\tht)$ with a Gaussian distribution. The Laplace approximation is obtained by performing a second-order Taylor series expansion about the mode of $\log\{p(\w|\y,\tht)\}$, resulting in a Gaussian approximation with mean equal to the mode and precision matrix equal to the Hessian at the mode.  Appendix \ref{la} provides more details on general Laplace approximations.    

We obtain a Laplace approximation of the form, 
\[\w|\y, \tht^{(t)} \sim N\!\left(\w^{(t)}, (\Sigma^{-1}_{\et^{(t)}}+C_{\w^{(t)}})^{-1}\right),\]
where $\w^{(t)}$ is the mode of $p(\w|\y, \tht^{(t)})$, 
$C_{\w^{(t)}}$ is a diagonal matrix with diagonal entries $\dt\, \circ\, \exp(\w^{(t)})$, and $\dt$ represents the $n\times1$ vector of pixel areas. To obtain $\w^{(t)}$, we select a starting value $\w^{(t)}_0 = \w^{(t-1)}$, and then iterate as
\begin{align}\label{newton_update}
\w^{(t)}_{\ell+1} = 
\w^{(t)}_{\ell} + 
\left(\Sigma^{-1}_{\et^{(t)}} + C_{\w^{(t)}_{\ell}}\right)^{-1}
\left(\y - \dt\circ\exp(\w^{(t)}_{\ell}) - \Sigma^{-1}_{\et^{(t)}}(\w^{(t)}_{\ell}- X \bt^{(t)})\right),
\end{align}
which corresponds to performing Newton-Raphson updates. Convergence is determined by $n^{-1/2} \| \w_{\ell+1}^{(t)} - \w_{\ell}^{(t)} \| <\epsilon$, where we use $\epsilon$ = $1\times 10^{-3}$.  We set $\w^{(t)}$ equal to the converged value.  

We use preconditioned conjugate gradient (PCG) to solve the system of equations required to evaluate \eqref{newton_update} \citep{hestenes1952methods}. PCG is a standard algorithm for solving positive definite systems. In this case, PCG relies on successive matrix-vector multiplication involving  $\Sigma^{-1}_{\et^{(t)}} + C_{\w^{(t)}_{\ell}}$ and a preconditioning matrix. This is computationally efficient because $\Sigma^{-1}_{\et^{(t)}}$ is block circulant, so the multiplication can be done with FFTs, and  $C_{\w^{(t)}_{\ell}}$ and our preconditioning matrix are diagonal. The PCG algorithm is included in Algorithm 1 of Appendix \ref{pcg}. 
Having obtained the Laplace approximation, we approximate the posterior mean $E_{\w|\y,\tht^{(t)}}[\,\w \,]$ with $\w^{(t)}$ and the posterior variance $E_{\w|\y,\tht^{(t)}}[\, (\w - E_{\w|\y,\tht^{(t)}}[\,\w \,])(\w - E_{\w|\y,\tht^{(t)}}[\,\w \,])^T\,]$ with $(\Sigma^{-1}_{\et^{(t)}}+C_{\w^{(t)}})^{-1}$.

\subsection{Hutchinson Trace Approximation}

The Laplace approximation allows us to replace  $E_{\y,\tht^{(t)}}[\, (\w - E_{\y,\tht^{(t)}}[\,\w \,])(\w - E_{\y,\tht^{(t)}}[\,\w \,])^T\,]$ with $(\Sigma^{-1}_{\et^{(t)}}+C^{(t)})^{-1}$. Because $\textup{tr}\!\left(\Sigma^{-1}_{\et}(\Sigma^{-1}_{\et^{(t)}}+C^{(t)})^{-1}\right)$  remains challenging to evaluate,
we use the Hutchinson trace approximation (HTA) \citep{hutchinson1989stochastic}. HTA is a technique for calculating the trace when a matrix $A$ is too hard to compute, but performing matrix-vector multiplication, $\mathbf{v}^{T}A\mathbf{v}$, is feasible.  In this context, $A = \Sigma^{-1}_{\et}(\Sigma^{-1}_{\et^{(t)}}+C^{(t)})^{-1}$. Feasibility of evaluating $\mathbf{v}^{T}A\mathbf{v}$ is a consequence of the assumed circulant structure of $\Sigma^{-1}_{\et}$ and the relationship between evaluation of $(\Sigma^{-1}_{\et^{(t)}}+C^{(t)})^{-1}\mathbf{v}$ and evaluation of the Newton-Raphson updates used to compute $\w^{(t)}$.
For $M \geq 1$ random vectors $\mathbf{v}_i$ with independent, identically distributed Rademacher distributed elements, HTA approximates $\textup{tr}\!\left(\Sigma^{-1}_{\et}(\Sigma^{-1}_{\et^{(t)}}+C^{*})^{-1}\right)$ with
\[\frac{1}{M}\sum_{i = 1}^M\mathbf{v}_{i}^{T}\left(\Sigma^{-1}_{\et}(\Sigma^{-1}_{\et^{(t)}}+C^{(t)})^{-1}\right)\mathbf{v}_{i}.\]
For each $\mathbf{v}_i$, we use the PCG algorithm described in Section \ref{subsubsec:lapapp} to quickly solve $(\Sigma^{-1}_{\et^{(t)}}+C^{*})\mathbf{r}_i = \mathbf{v}_{i}$. We then leverage FFTs to efficiently evaluate $\frac{1}{M}\sum_{i = 1}^M\mathbf{v}_{i}^{T}\Sigma^{-1}_{\et}\mathbf{r}_i$.
Smaller values of $M$ yield faster, but less accurate, approximations to $\textup{tr}\!\left(\Sigma^{-1}_{\et}(\Sigma^{-1}_{\et^{(t)}}+C^{(t)})^{-1}\right)$. 

\subsection{Defining and Increasing the Approximate Objective Function}\label{mstep}

We define an approximate objective function $\widetilde{Q}(\tht|\tht^{(t)}; M)$, via combining the Laplace and Hutchinson trace approximations,
\begin{align}
\widetilde{Q}(\tht|\tht^{(t)}; M) &= -\frac{1}{2}\bigg[\log(|\Sigma_{\et}|)+ 
(\w^{(t)} - X\bt)^T\Sigma^{-1}_{\et}(\w^{(t)} - X\bt) \label{eq:objfun} \\ 
&+ \frac{1}{M}\sum_{i = 1}^M\mathbf{v}_{i}^{T}\left(\Sigma^{-1}_{\et}(\Sigma^{-1}_{\et^{(t)}}+C^{(t)})^{-1}\right)\mathbf{v}_{i}\bigg]. \nonumber
\end{align}
We consider the problem of finding a new value $\tht^{(t+1)}$ that satisfies $\widetilde{Q}(\tht^{(t+1)}|\tht^{(t)}; M) > \widetilde{Q}(\tht^{(t)}|\tht^{(t)}; M)$. First, we set
\begin{align*}
\bt^{(t+1)}&= \underset{\bt}{\mathrm{argmax}}(\w^{(t)} - X\bt)^T\Sigma^{-1}_{\et^{(t)}}(\w^{(t)} - X\bt).
\end{align*}
This is equivalent to computing the regression coefficients for a regression of the mode $\w^{(t)}$ on the predictors with error covariance $\Sigma_{\et^{(t)}}$. 

Next, we set
\begin{align*}
\et^{(t+1)}= \underset{\et}{\mathrm{argmax}}&-\frac{1}{2}\left[\log(|\Sigma_{\et}|)+ 
(\w^{(t)} - X\bt^{(t+1)})^T\Sigma^{-1}_{\et}(\w^{(t)} - X\bt^{(t+1)})\right.\label{eq:objfun} \\ 
&+\left.\frac{1}{M}\sum_{i = 1}^M\mathbf{v}_{i}^{T}\left(\Sigma^{-1}_{\et}(\Sigma^{-1}_{\et^{(t)}}+C^{(t)})^{-1}\right)\mathbf{v}_{i}\right].
\end{align*}
This yields a new value $\tht^{(t+1)} = (\bt^{(t+1)}, \et^{(t+1)})$ that satisfies $\widetilde{Q}(\tht^{(t+1)}|\tht^{(t)}; M) \geq \widetilde{Q}(\tht^{(t)}|\tht^{(t)}; M)$.

\subsection{Practical Implementation Details}\label{deets}

We iterate between the E- and M-steps until we reach convergence, or reach the maximum number of user-specified iterations, whichever comes first.  Our convergence criterion is 
\[\sqrt{\frac{1}{p + 3}\sum_{i = 1}^{p+3}\left(\tht_i^{(t+1)}-\tht_i^{(t)}\right)^2}<\epsilon,\]
where $p + 3$ corresponds to the length of $\tht$ and $\epsilon =  10^{-5}$.  Given the approximations throughout this method, convergence is not guaranteed.  However, simulations suggest that iterating through about 100 EM steps provides reasonable results, even if the algorithm has not converged, so we set the maximum number of iterations to 100.

Implementation of this EM algorithm requires specification of starting values $\tht^{(0)} = (\bt^{(0)}, \et^{(0)})$. We recommend setting $\bt^{(0)} = \boldsymbol 0$ and $\et^{(0)} = \widetilde{\et}^*$ where $\widetilde{\et}^*$ refers to the EM estimate of variance parameters based on assuming an LGCP model with no predictors for the same data. For EM estimation of the variance parameters under an LGCP model with no predictors, we recommend the initial value $\widetilde{\tht}^{(0)} = (\widetilde{\et}^{(0)}) = (\bar{Y},  n_1/4 )$.

We refer to the implementation scheme described above as the ``joint" implementation because we update both $\bt$ and $\et$ during each M-step. We also consider the alternative method of fixing $\bt$ at the generalized least squares estimate and 
updating only $\et$ at each $M$-step. We will refer to our implementation of this alternative scheme as the ``fixed" case.
  
\subsection{Recovery of the Residual Latent Field}

Having obtained an optimal value of the parameters $\boldsymbol{\tht}^* = (\bt^*, \et^*)$, we can recover $\z^*$, the posterior mode of the latent field $\z$ as defined in the original stochastic representation of the LGCP model, from $\w^*$, the posterior mode of $\w$ at $\tht^*$, by setting $\z^* = \w^* - X \bt^*$.  A detailed derivation is provided in Appendix \ref{supp_recovery}.  

In practice, it can also be of interest to approximate the posterior mean of the latent field on the intensity scale, $E_{\y, \tht^*}[\,\text{exp}(\z) \,]$, as opposed to the log scale.
Again, we use a Laplace approximation to the posterior distribution to approximate this expectation. The Laplace approximation to the posterior distribution of $\z$ given $\y$ and $\tht^*$ is
\[\z|\y, \tht^{*} \sim N(\z^{*}, \boldsymbol{\Psi}^{-1} ),
\]
where $\boldsymbol{\Psi} = \Sigma^{-1}_{\et^{*}}+\textup{diag}(\dt\circ\exp(X\bt^* + \z^{*}))$.
Given the Laplace approximation for $\z|\y, \tht^{*}$, it follows that $\exp(\z)|\y,\tht^{*}$ has a multivariate log-normal distribution with 
\begin{align*}
E[\,\exp(Z_{j})|\y,\tht^{*}\,] &\approx \exp\left(Z^{*}_j+\frac{1}{2}\boldsymbol{\Psi}^{-1}_{jj}\right).
\end{align*}
It is too computationally expensive to invert the dense $n\times n$ matrix $\boldsymbol{\Psi}$ and extract the diagonal elements. Instead, we propose a local approximation.
For each $j$, we extract the entries of $\boldsymbol{\Psi}$ corresponding to the $k \times k$ square neighborhood of pixels surrounding pixel $j$. We then invert the submatrix containing these entries and extract the diagonal entry of the inverse corresponding to pixel $j$ as our approximation to $\boldsymbol{\Psi}^{-1}_{jj}$.
Note that the circulant covariance structure has an implied assumption that $\z$ is dependent across opposite boundaries of the domain.  Likewise, constructing neighborhood submatrices for pixels along one edge of the spatial domain involves incorporating pixels from the other edge of the spatial domain.
Like the value of $M$ used to construct the HTA, smaller values of $k$ yield faster, but less accurate, approximations to $E[\,\exp(\z)|\y,\tht^{*}\,]$

\section{Simulation Study}\label{sim_res}

Because our initial exploration of the data in Section~\ref{prelim_light} suggested that VL is the most competitive alternative method, we focus on SLEM and VL in simulations.  For each method, we implement both the fixed and joint implementations suggested in Section \ref{deets}. For SLEM, we vary $M$, the number of vectors in the HTA.  We compare average runtime, estimates of $\bt$, and average root-mean-square-error of the log-intensity $\text{log}(\boldsymbol \lambda)$ across $100$ simulation replicates of $\y$, corresponding to a $70\times 70$ grid. 

Each simulation replicate uses the same $\z$, and thus the same intensity.  We simulate $\z$ from a multivariate normal distribution with zero mean and Mat\'ern covariance with variance $\sigma^2 = 2$, range $\alpha = 18$, and smoothness $\nu = 1$.  We define $\boldsymbol X(s_i)$ to include an intercept and several covariates, and consider two different settings for covariate construction. The first covariate setting produces a noisy true intensity using three covariates, where two covariates are simulated from a standard normal distribution and another is constructed from raw Channel 5 ABI data.  The second covariate setting produces a smooth true intensity using two covariates, where the two covariates are constructed from raw Channel 5 and Channel 8 ABI data, respectively. We refer to the first noisy true intensity setting as ``Setting 1" and the second smooth true intensity setting as ``Setting 2."

\begin{table}
\begin{center}
\resizebox{\textwidth}{!}{%
\begingroup\small
\begin{tabular}{ccccccccc}
 {\bfseries Method} & {\bfseries Update} & {\bfseries M} & {\bfseries Time (min.)} & {\bfseries $\beta_0$} & {\bfseries $\beta_1$} & {\bfseries $\beta_2$} & {\bfseries $\beta_3$} & {\bfseries RMSE(log($\lambda$))} \\ 
  \hline
SLEM & fixed & 1 & 3.71 & 1.98 (0.98) & 0.8 (0.05) & 0.6 ($<$0.01) & 0.68 (0.27) & 0.279$|$0.19 \\ 
  SLEM & fixed & 10 & 21.24 & 1.98 (0.98) & 0.8 (0.05) & 0.6 ($<$0.01) & 0.68 (0.27) & 0.279$|$0.19 \\ 
  SLEM & joint & 1 & 14.21 & 1.05 (0.05) & 0.85 ($<$0.01) & 0.6 ($<$0.01) & 0.88 (0.07) & 0.269$|$0.171 \\ 
  SLEM & joint & 10 & 71.43 & 1.05 (0.05) & 0.85 ($<$0.01) & 0.6 ($<$0.01) & 0.88 (0.07) & 0.269$|$0.171 \\ 
  VL & fixed & - & 5.26 & 1.98 (0.98) & 0.8 (0.05) & 0.6 ($<$0.01) & 0.68 (0.27) & 0.178$|$0.178 \\ 
  VL & joint & - & 49.46 & 1.13 (0.15) & 0.85 (0.01) & 0.6 ($<$0.01) & 0.94 (0.01) & 0.137$|$0.136 \\ 
  \end{tabular}
\endgroup
}
\caption{Results of applying SLEM and VL to Setting 1.  True $\bt= 1, 0.85, 0.6, 0.95$.  Parameter estimates reported as Mean (RMSE), where RMSE stands for `root-mean-square-error' throughout.  Runtime presented as average over 100 trials.  RMSE(log$\lambda$) reported as ``full grid$|$interior points", where the interior point calculation is restricted to to pixels $3:(n_1-2) \times 3:(n_1-2)$, i.e. 2 pixels in from the edge of the spatial domain.}
\label{tab:sim1_res}
\end{center}
\end{table} 

\begin{figure}
\begin{center}
\includegraphics[width = \textwidth]{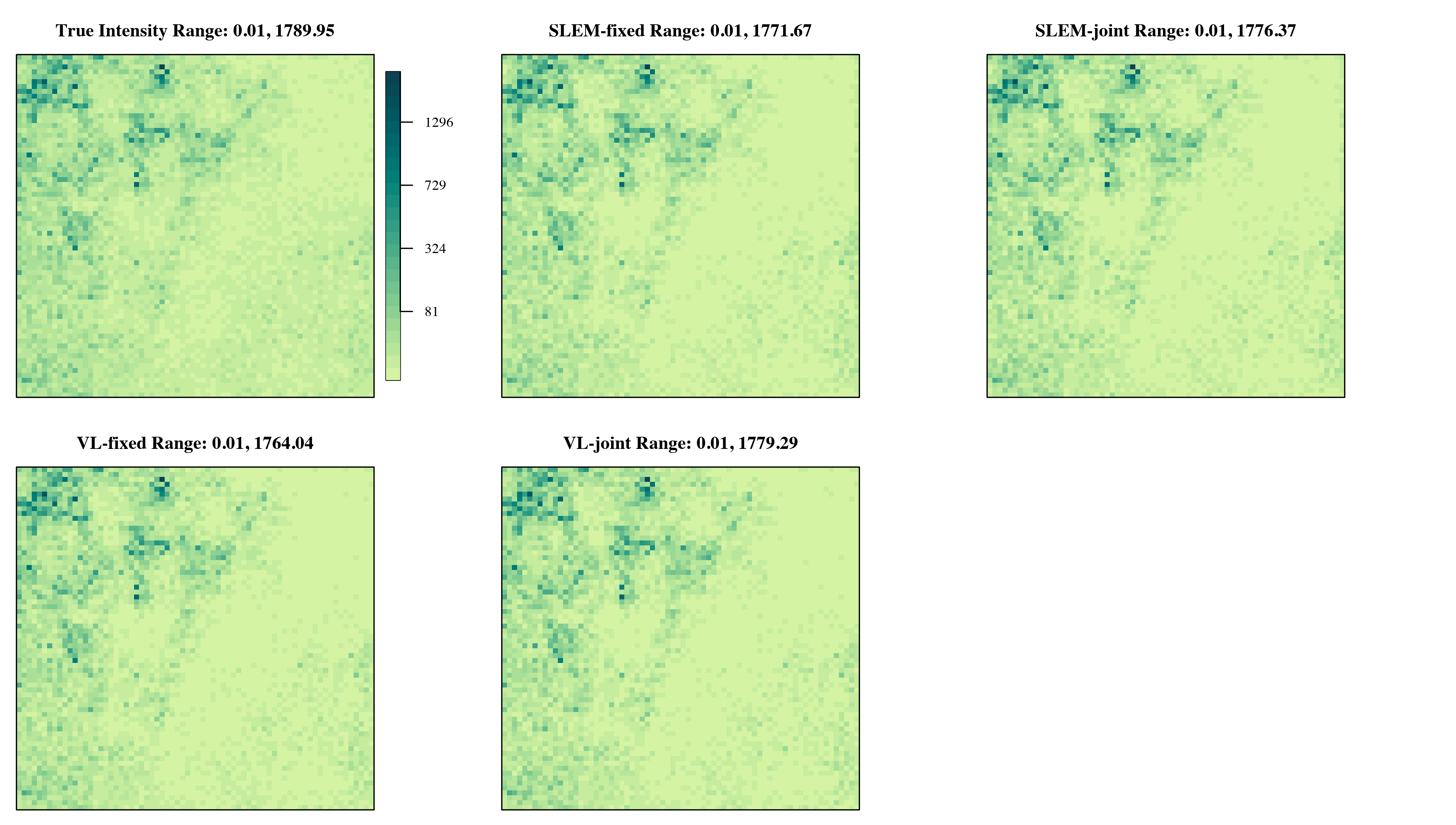}
\end{center}
\caption{Examples of intensities resulting from applying SLEM with $M = 1$ and VL to Setting 1.}
\label{fig:sim1_res}
\end{figure}

The results for Setting 1 are summarized in Table \ref{tab:sim1_res} and Figure \ref{fig:sim1_res}.  In SLEM, we see that increasing $M$ affects  runtime but not estimation of $\bt$ or the  log-intensity. In general, the joint settings are slower but more accurate in terms of RMSE than their fixed counterparts. SLEM-joint is more accurate than VL-joint for estimating $\beta_0$, but VL-joint is more accurate than SLEM-joint for estimating $\beta_3$ and the log-intensity. However, VL-joint takes three times as long as SLEM-joint with $M=1$ to run.

Figure \ref{fig:sim1_res} shows that all of the methods produce visually similar intensity estimates. SLEM-joint sacrifices some accuracy along the boundary of the domain, which is not surprising since circulant covariance methods are known to suffer from edge effects. This is also reflected in Table \ref{tab:sim1_res}, which includes RMSE for the log-intensity restricted to interior points. The accuracy advantage of VL for estimating the log-intensity shrinks when only interior points are considered, although the VL-joint estimates remain superior.  The same  conclusions are echoed in the analysis of Setting 2, which is described in Appendix C.

\section{Lightning Data}\label{real_res}

\subsection{Covariate Construction and Selection}\label{cov_descr}

For the lightning data, we define $\boldsymbol X(s_i)$ to include an intercept, two covariates derived from ABI data, and a third covariate measuring elevation. All covariates are centered and scaled to facilitate comparison of parameter estimates. 

The first two covariates ($X_1$ and $X_2$) are constructed from brightness temperature from ABI Channel 13, which serves as a proxy for cloud-top temperature \citep{henderson2021evaluating}. Channel 13 measures light at an infrared wavelength (10.3 $\mu m$), which ensures continual measures throughout day and night.  Meteorologists use cloud-top temperature, which is inversely related with cloud-top height, to monitor updrafts in severe convective storms \citep{mecikalski2006forecasting}. Some care must be taken to construct covariates capable of connecting the minute-by-minute ABI data to lightning counts, which we have aggregated over one-hour periods. Simple averaging of the Channel 13 data over one hour may not capture the dynamic and transitory nature of clouds in severe storms. \cite{lee2021simplified} suggested constructing variables based on 10 minute intervals of data. 
Moreover, both the absolute cloud heights and sharp changes in cloud heights are important factors impacting the severity of storms.

For these reasons, we construct $X_1$ and $X_2$ as follows.
For pixel $i$ and minute $t$, let $A_{i,t}$ be the Channel 13 brightness temperature.  Letting $k = 1, \dots, 6$, we define six 10-minute proxies for differenced and absolute cloud top temperatures for one hour of data as follows:
\begin{align*}
X_{1,k} &= A_{i,10k} - A_{i,10k-9} &&(\mbox{10 minute differences}), \\
X_{2,k} &= \frac{1}{10}\sum_{j=1}^{10} A_{i,10k + j} &&(\mbox{10 minute averages)}.
\end{align*}
The differences are designed to capture changes in cloud top height, whereas the averages measure absolute cloud top height.
We next consider several functions of these six values: the average, minimum, maximum, and range. For both differences and averages, the best function of the six values is determined by selecting the function which produces the highest log-likelihood of a simple Poisson regression with a single covariate. Through this process we select the average of the 10 minute differences and the minimum of the 10-minute averages as our covariates:
\begin{align*}
X_1 &= \frac{1}{6} \sum_{k=1}^6 X_{1,k}, \\
X_2 &= \min\{ X_{2,1},\ldots,X_{2,6} \}.
\end{align*}
Note that the inverse relationship between cloud-top temperature and cloud-top height allows us to interpret the minimum of Channel 13 brightness temperatures ($X_2$) as a proxy for the maximum cloud-top height.  We interpret the average of Channel 13 brightness temperature differences ($X_1$) as a proxy for cloud growth \citep{henderson2021evaluating}.

We also include elevation ($X_3$) as an environmental factor in our model, because a connection between lightning and elevation has been hypothesized in other parts of the world \citep{kilinc2007spatial, kotroni2008lightning}. Elevation data is available from the ETOPO5 data repository, which contains  land and sea-floor elevation at an approximate 8 km resolution over the United States \citep{noaa1988data}. We assemble the $125\times125$ grid of elevation data in each hour by matching each pixel in the 125$\times$125 grid used for the lightning and ABI data to the locations in the ETOPO5 dataset.

\subsection{Analysis}
We now analyze the lightning data from Figures \ref{fig:real1} and \ref{fig:real2} which motivated this work. Based on the results from the simulation study, we only consider the results from the joint implementation of both VL and SLEM, as this scheme yielded the most accurate results for each method. To facilitate out-of-sample comparisons, we fit each model to a random subset of 90\% of the strikes and test the resultant model on the remaining 10\% of the strikes. Note that we subsample the strikes, not the pixels, so we always work on a full grid with no missing values.   We record the log-score, or log-likelihood value associated with the testing data, given a 10/90 scaling of the fitted intensity.  This corresponds to evaluating
\[ \mbox{log-score} = \sum_{i=1}^n Y_i * \left(\log\Delta_i + \log\lambda_i + \log(1/9)\right) - \left((1/9) * \Delta_i * \lambda_i\right) - \log Y_i!\]
on the testing data values $Y_i$ and estimated intensities $\lambda_i$ from the training data.  We also offer visuals of the estimated intensity functions for qualitative comparisons.

\begin{table}
\begin{center}
\begingroup\small
\begin{tabular}{ccccccc}
 {Method} & {Time (min.)} & {Log Score} & {$\beta_0$} & {$\beta_{1}$ (avg diff)} & {$\beta_{2}$ (min avg)} & {$\beta_{3}$ (elev)} \\ 
  \hline
SLEM & \textbf{549} & \textbf{-1853} & -10.06 & -0.12 & -1.13  & 0.4 \\ 
  VL & 1670 & -2265 & -7.72 & -0.29 & -3.39 & 0.1 \\ 
  \end{tabular}
\endgroup

\caption{Results for Lightning Dataset 1.}
\label{tab:real_res1}
\end{center}
\end{table} 

\begin{table}[t!]
\begin{center}
\begingroup\small
\begin{tabular}{ccccccc}
 {Method} & {Time (min.)} & {Log Score} & {$\beta_0$} & {$\beta_{1}$ (avg diff)} & {$\beta_{2}$ (min avg)} & {$\beta_{3}$ (elev)} \\ 
  \hline
SLEM & \textbf{35} & \textbf{-2004} & -8.46 & 0.10 & -0.75 & 0.55 \\ 
  VL & 854 & -2541 & -5.58 & -0.15 & -1.01 & 0.29 \\ 
  \end{tabular}
\endgroup

\caption{Results for Lightning Dataset 2.}
\label{tab:real_res2}
\end{center}
\end{table}

\begin{figure}[h!]
\begin{center}
\includegraphics[width = \textwidth]{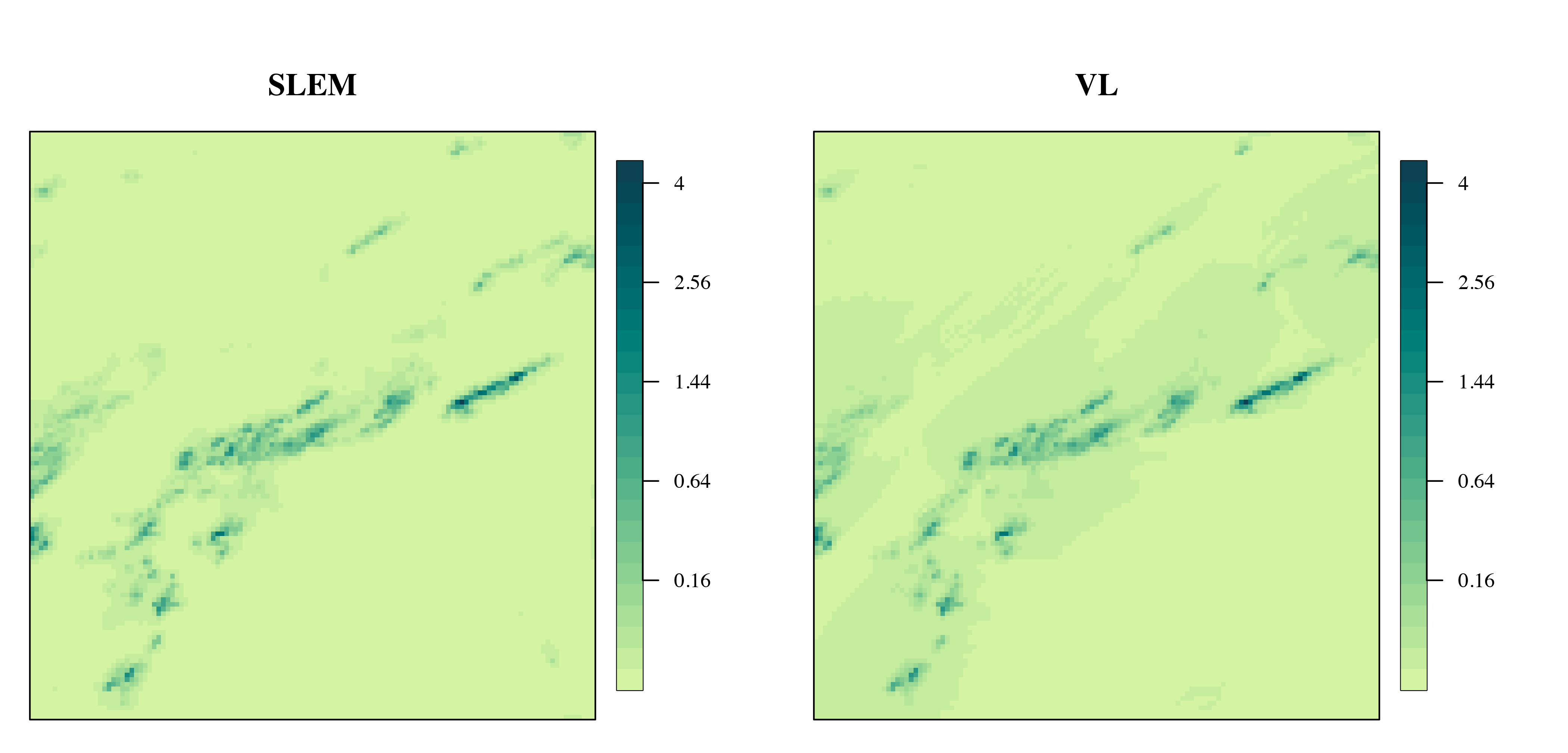}
\end{center}
\caption{Estimated intensities/1 km$^2$/hour resulting from the SLEM and VL algorithms applied to lightning dataset 1 in testing regime (see Figure~\ref{fig:real1}).  We employ a square-root transformation for better image definition.}
\label{fig:real1_testing}
\end{figure}

\begin{figure}[t!]
\begin{center}
\includegraphics[width = \textwidth]{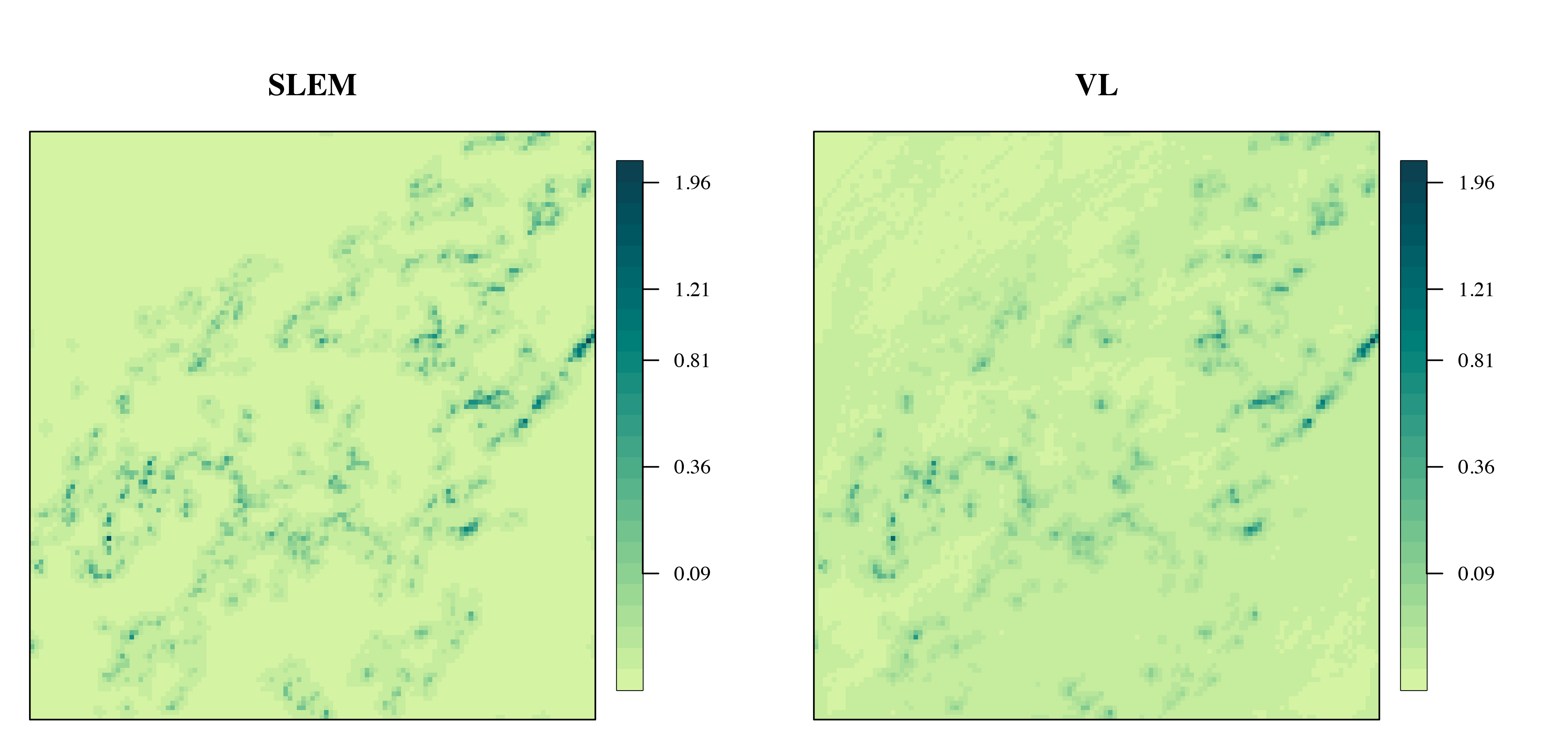}
\end{center}
\caption{Estimated intensities per km$^2$/hour resulting from the SLEM and VL algorithms applied to lightning dataset 2 in testing regime (see Figure~\ref{fig:real2}). We employ a square-root transformation for better image definition.}
\label{fig:real2_testing}
\end{figure}

As seen in Tables \ref{tab:real_res1} and \ref{tab:real_res2}, SLEM boasts the largest log-score on test data, indicating a superior model fit.  VL's lower log-scores are likely a result of the overly smooth estimates of the intensities, as seen in Figures \ref{fig:real1_testing} and \ref{fig:real2_testing}, which fail to accurately reflect the isolated regions of lightning activity present in the data.  In contrast, SLEM appears, both quantitatively and qualitatively, to capture both the isolated regions of lightning activity and those larger areas with sufficient detail, especially along the interior points which are not affected by edge effects.  

Turning our attention to the estimated regression coefficients, we see that for both datasets and for both estimation methods, the minimum of the 10 minute average of brightness temperatures has the strongest effect. The effect is always negative, which is expected, since the minimum of the average brightness temperatures is negatively associated with the maximum of the average cloud-top heights, and high cloud tops indicate severe weather. Elevation has a smaller, but positive effect. The weakest effect is the average of the 10 minute differences, which mostly have small negative effects. Negative effects are also expected, since negative differences correspond to cloud growth. For both datasets, VL puts more weight on the minimum average than SLEM, and less weight on elevation. SLEM runs much faster than VL--about 3 times faster on Dataset 1 and more than 20 times faster on Dataset 2.

\section{Discussion}\label{conc}

In this paper, we introduced SLEM, a new approximate method for fitting LGCP models to large spatial point pattern datasets.  This method leveraged spectral, Laplace, and Hutchinson trace approximations  to make computational gains without sacrificing much accuracy.  We verified this in a simulation study where SLEM ran much faster than VL and competitively estimated $\boldsymbol \beta$.  Via simulations, we also showed SLEM is robust to the choice of  $M$, the number of vectors in the HTA.  When applied to the lightning data, SLEM was superior to its chief competitor, running faster and with more accuracy in both examples. 
Moreover, our application to the lightning data produces findings that are consistent with the meteorology literature. Specifically, we found that high and rapidly increasing cloud tops over high elevation regions are associated with more lightning strikes. 

Although we developed SLEM in the context of the lightning data described in this paper and focus on lightning data throughout, SLEM could be applied more generally to large spatial point pattern datasets, which are ubiquitous in a variety of fields including epidemiology and finance.  SLEM acts as a new tool not only for researchers seeking to understand lightning dynamics, but also for those looking to model and investigate large spatial point pattern datasets, in general. 

We conclude by describing several potential extensions. 
First, this work could be extended to accommodate spatio-temporal data and used to model the more complex relationships between space, time, and the covariates in question. The computational advantages offered by SLEM would be especially valuable for spatio-temporal models due to the large size of spatio-temporal data.  Second, existing methods for reducing edge effects could be incorporated into SLEM. In particular, methods that embed the circulant covariance on a larger spatial domain could be adapted for use with SLEM in order to gain an edge-effect free approximation of those covariances within the spatial domain of interest \citep{guinness2017Circulant}.

\section{Acknowledgements}
The authors would like to thank Finn Lindgren and Matthias Katzfuss for their aid in implementing INLA and VL, respectively.  
The authors gratefully acknowledge financial support from the National Science Foundation 1455172, 1916208, 1934985, 1940124, 1940276, 1953088, and 2114143, USAID 7200AA18CA00014, National Institutes of Health  R01ES027892, and Cornell Atkinson Center for Sustainability.

\newpage

\appendix
\section{Additional Component Techniques for SLEM}

\subsection{Laplace Approximation}\label{la}
Consider a random variable, $\mathbf{w} = (w_1,\hdots,w_n)$, whose density has the form, 

\[p(\mathbf{w}) \propto \exp\left\{-\frac{1}{2}\mathbf{w}^{\top}Q\mathbf{w} + \sum_{i\in\mathcal{I}}g_i(w_i)\right\}.\]

The Laplace approximation is obtained by performing a second-order Taylor series expansion at the mode of the distribution in question, resulting in an Gaussian approximation with mean equal to the mode and precision matrix equal to the Hessian at the mode.  To make this more concrete, consider the mode estimate $\mu^{(t)}$ at iteration $t$.  The second order expansion of $g_i(w_i)$ around $\mu^{(t)}$ is 

\[g_i(w_i) \approx g_i(\mu^{(t)}_i) + b_iw_i - \frac{1}{2}c_iw_i^2,\]

\noindent
where $b_i$ and $c_i$ depend on $\mu^{(t)}$.  The Gaussian approximation at iteration $t+1$ has a mean equal to the mode, $\mu^{(t+1)}$ -- the solution to $\{Q + \textup{diag}({\mathbf{c}_{\mu^{(t)}}})\}\mu^{(t+1)} = \mathbf{b}$ -- and precision matrix $Q + \textup{diag}(\mathbf{c}_{\mu^{(t+1)}})$.  This iterative computation of the mode, and corresponding precision matrix, is performed via a Newton-Raphson method, and continues until reaching some threshold of convergence.  For particularly quick evaluation, one can solve $\{Q + \textup{diag}({\mathbf{c}_{\mu^{(t)}}})\}\mu^{(t+1)} = \mathbf{b}$ via a preconditioned conjugate gradient scheme.  We adopt this in SLEM.  The specific convergence criterion we use is 
\[\sqrt{\frac{1}{n}\sum_{i = 1}^{n}\left(\mu_i^{(t)}-\mu_i^{(t-1)}\right)^2}<\epsilon.\]  

\subsection{Preconditioned Conjugate Gradient}\label{pcg}

Conjugate gradient is a technique for solving a system of linear equations, i.e.\ efficiently solving $A\mathbf{x} = \mathbf{b}$, where A is a $n\times n$ symmetric positive definite matrix \citep{hestenes1952methods}.  The preconditioned variant introduces a preconditioner matrix $M$ such that $M^{-1}A$ has a smaller condition number than $A$, thus leading to faster convergence.  Several standard preconditioner matrices include,  
\begin{itemize}
\item Jacobi(diagonal): $M = \textup{diag}(A)$,
\item Gauss-Seidel: $M = \textup{diag}(A) + L_{A}$; $L_{A}$ strictly lower diagonal part of $A$,
\item Successive over-relaxation: $M = \frac{1}{\omega}\left(\textup{diag}(A) + \omega L_{A}\right)$; $0<\omega<2$.  
\end{itemize} 
\noindent We use the Jacobi preconditioner in this work.  The exact algorithm for the preconditioned conjugate gradient method is stated below.

\begin{algorithm}
\caption{Preconditioned Conjugate Gradient Method for Solving the Symmetric Positive Definite System $A\mathbf{x} = \mathbf{b}$,}
\begin{algorithmic}
\STATE Input: starting $x^{(0)}$ and stopping criterion $\epsilon$ 
\STATE Set: $k=0$, $\mathbf{r}^{(k)}=\mathbf{b}-A\mathbf{x}^{(k)}$, 
     $\mathbf{z}^{(k)}=M^{-1}\mathbf{r}^{(k)}$, $\mathbf{p}^{(k)} = \mathbf{z}^{(k)}$
\WHILE{$\sqrt{\frac{1}{n}\sum_{i=1}^n(\mathbf{r}_i^{(k+1)} - \mathbf{r}_i^{(k)})^2} > \epsilon$}{
\STATE $\alpha_k = \frac{\mathbf{r}^{(k)\mathrm{T}}\mathbf{z}^{(k)}}{\mathbf{p}^{(k)\mathrm{T}}A\mathbf{p}^{(k)}}$
\STATE $\mathbf{x}^{(k+1)} = \mathbf{x}^{(k)} + \alpha^{(k)}\mathbf{p}^{(k)}$
\STATE $\mathbf{r}^{(k+1)} = \mathbf{r}^{(k)} - \alpha^{(k)}A\mathbf{p}^{(k)}$
\STATE $\mathbf{z}^{(k+1)} = M^{-1}\mathbf{r}^{(k+1)}$
\STATE $\beta^{(k)} = \frac{\mathbf{z}^{(k+1)\mathrm{T}}\mathbf{r}^{(k+1)}}{\mathbf{z}^{(k)\mathrm{T}}\mathbf{r}^{(k)}}$ 
\STATE $\mathbf{p}^{(k+1)} = \mathbf{z}^{(k+1)} + \beta^{(k)}\mathbf{p}^{(k)}$
\STATE $k = k + 1$
}
\ENDWHILE
\RETURN $\mathbf{x}^{(k)}$
\end{algorithmic}
\end{algorithm}

Our implementation of PCG uses a tolerance of $\epsilon = 1\times 10^{-3}$.

\section{Recovery of the Residual Latent Field}\label{supp_recovery}

As the mode of the posterior distribution of $\w$ given $\y$ and $\tht^*$, $\w^*$, satisfies
\begin{align}\label{eq:wopt}
    \y - \dt \circ \text{exp}(\w^*) - \Sigma^{-1}_{\et^*} (\w^* - X\bt^*) = \boldsymbol 0,
\end{align}
where `$\circ$' refers to elementwise multiplication.
In contrast, the mode of the posterior distribution of $\w$ given $\y$ and $\tht^*$, $\z^*$, satisfies 
\begin{align}\label{eq:zopt}
    \y - \dt \circ \text{exp}(X \bt^* + \z^*) - \Sigma^{-1}_{\et^*} \z^*  = \boldsymbol 0.
\end{align}
Manipulating Equation~\eqref{eq:wopt} yields
\begin{align*}
    \y - \dt \circ \text{exp}(X\bt^* + \w^* - X \bt^*) - \Sigma^{-1}_{\et^*} (\w^* - X\bt^*) = \boldsymbol 0.
\end{align*}
Thus, $\z^* = \w^* - X \bt^*$ satisfies Equation~\eqref{eq:zopt} and the posterior mode $\z^*$ can be obtained by subtracting $X\bt^*$ from $\w^*$.

\newpage

\section{Supplemental Tables and Figures}\label{supp_figs}
\begin{figure}[h!]
\begin{center}
\includegraphics[width = \textwidth]{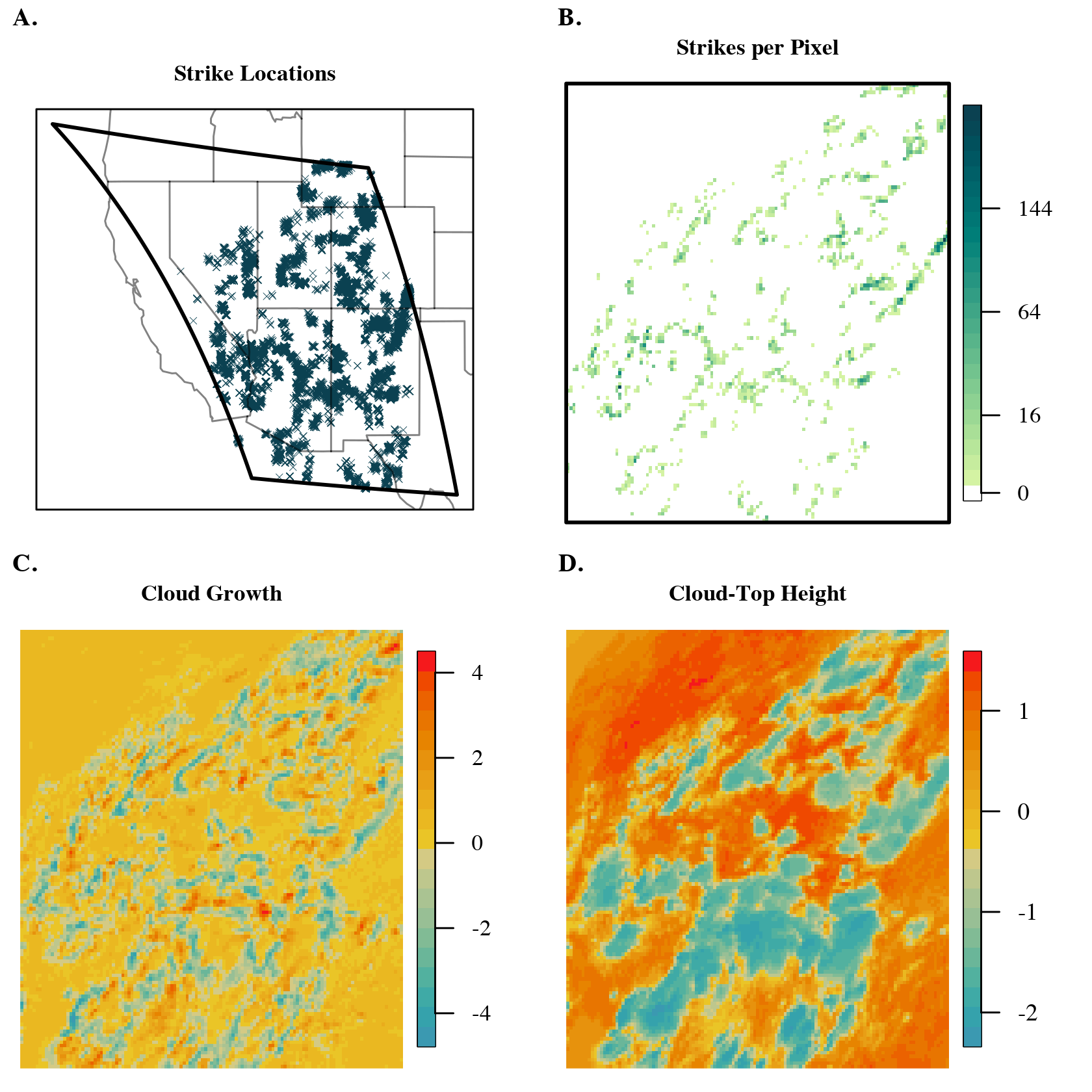}
\end{center}
\caption{\textbf{A.} Individual lightning strikes recorded from 20:00 - 20:59 GMT on 2018-07-11 in designated area of North America.  \textbf{B.} Lightning strikes from A. converted to counts per pixel on 125 x 125 grid.  We employ a square-root transformation for better image definition. \textbf{C. \& D.} Cloud growth and cloud-top height proxy data, respectively, averaged over same time frame as A. and grid as B.  Covariate values are centered and scaled here, and in the model.}
\label{fig:real2}
\end{figure}

\clearpage

\begin{figure}[h!]
\begin{center}
\includegraphics[width = \textwidth]{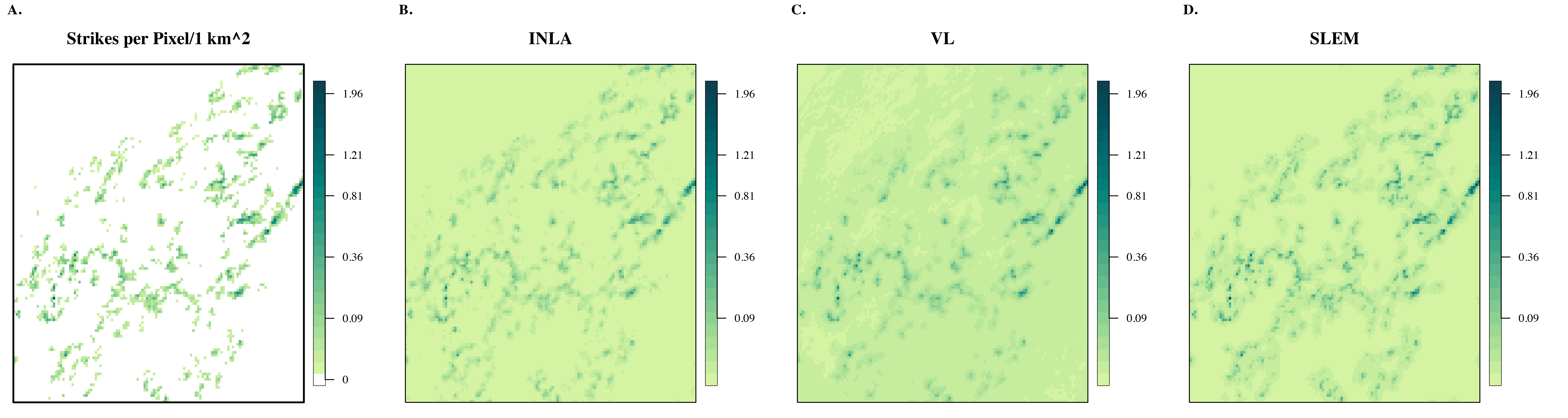}
\end{center}
\caption{Estimated intensity/1 km$^2$/hour returned by INLA (139 min.), VL (2247 min.), and SLEM (48 min.) fit to lightning dataset 2 (see Figure~\ref{fig:real2}).  We employ a square-root transformation for better image definition.}
\label{fig:real2_res_comp}
\end{figure}

\clearpage

\begin{table}[h!]
\begin{center}
\resizebox{\textwidth}{!}{%
\begingroup\small
\begin{tabular}{cccccccc}
 {\bfseries Method} & {\bfseries Update} & {\bfseries M} & {\bfseries Time (min.)} & {\bfseries $\beta_0$} & {\bfseries $\beta_1$} & {\bfseries $\beta_2$} & {\bfseries RMSE(log($\lambda$))} \\ 
  \hline
SLEM & fixed & 1 & 11.47 & 1.95 (0.7) & -1.44 (0.29) & -0.17 (0.67) & 0.406$|$0.321 \\ 
  SLEM & fixed & 10 & 75.48 & 1.95 (0.7) & -1.44 (0.29) & -0.17 (0.67) & 0.406$|$0.321 \\ 
  SLEM & joint & 1 & 28.23 & 1.3 (0.05) & -1.16 (0.02) & 0.46 (0.05) & 0.329$|$0.193 \\ 
  SLEM & joint & 10 & 160.48 & 1.3 (0.05) & -1.16 (0.02) & 0.46 (0.05) & 0.329$|$0.193 \\ 
  VL & fixed & - & 4.46 & 1.95 (0.7) & -1.44 (0.29) & -0.17 (0.67) & 0.328$|$0.34 \\ 
  VL & joint & - & 46.94 & 1.35 (0.14) & -1.14 (0.02) & 0.45 (0.06) & 0.164$|$0.162 \\ 
  \end{tabular}
\endgroup
}
\caption{Results of applying SLEM and VL to Setting 2.  True $\bt = 1.25, -1.15, .5$.  Parameter estimates reported as Mean (RMSE), where RMSE stands for `root-mean-square-error' throughout.  Runtime presented as average over 100 trials.  RMSE(log$\lambda$) reported as full grid|interior points, where the interior point calculation is restricted to to pixels $3:(n_1-2) \times 3:(n_1-2)$, i.e. 2 pixels in from the edge of the spatial domain.}
\label{tab:sim2_res}
\end{center}
\end{table} 

\begin{figure}[h!]
\begin{center}
\includegraphics[width = \textwidth]{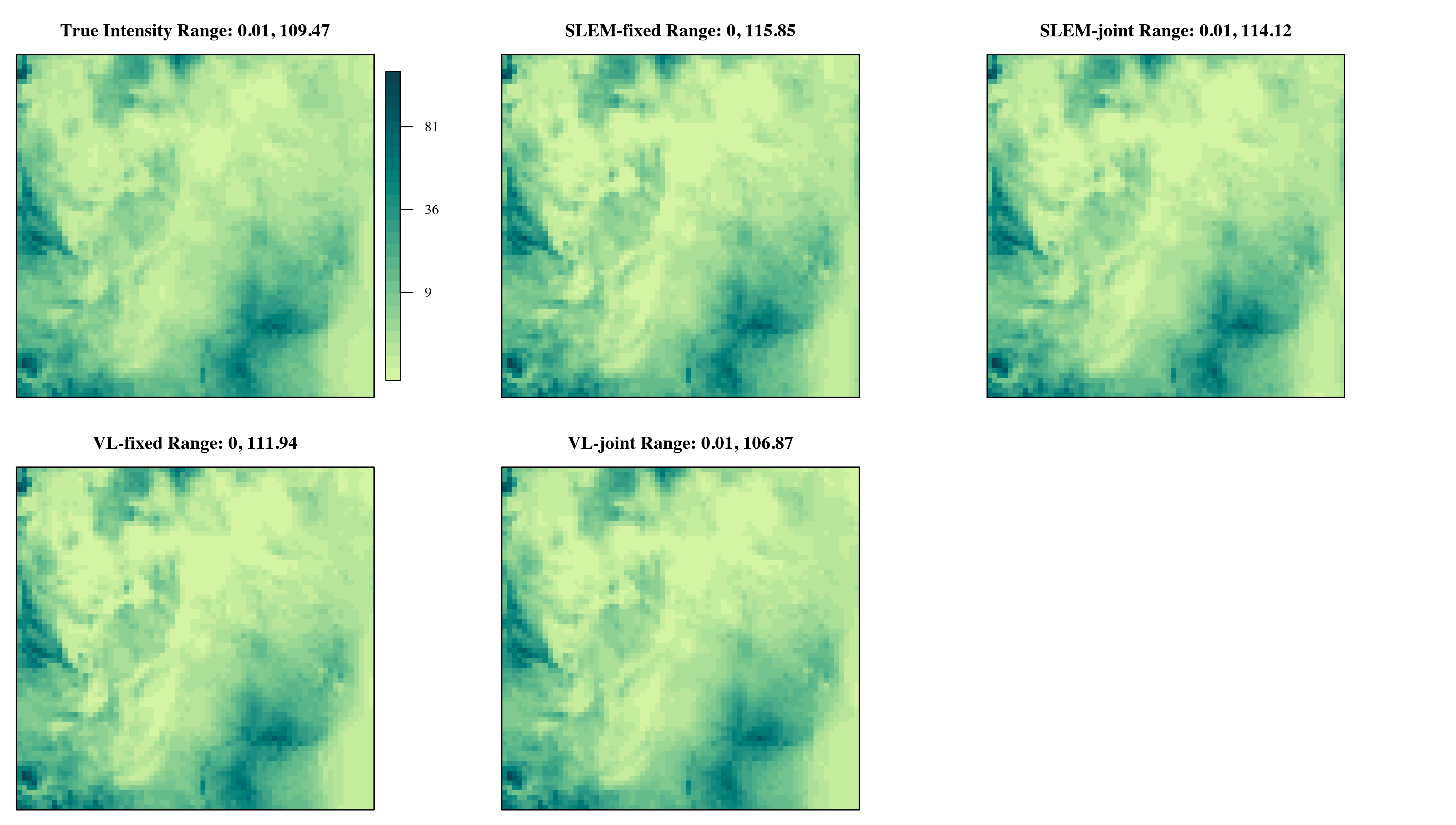}
\end{center}
\caption{Examples of intensities resulting from applying SLEM and VL to Setting 2.  SLEM results associated $M$= 1 - other settings not shown due to similarity.}
\label{fig:sim2_res}
\end{figure}

\clearpage

\bibliographystyle{plainnat}
\bibliography{light_bib}

\end{document}